\documentclass[twocolumn,amsmath,amssymb,aps,prb,floatfix,nofootinbib,superscriptaddress]{revtex4}
\usepackage{graphicx,graphics,times,color}
\usepackage{bm}
\usepackage{longtable}

\def\be{\begin{equation}}
\def\ee{\end{equation}}
\def\ba{\begin{eqnarray}}
\def\ea{\end{eqnarray}}
\def\la{\langle}
\def\ra{\rangle}

\begin{document}

\title{Kondo Cloud Mediated Long Range Entanglement After Local Quench in a Spin Chain }
\author{Pasquale Sodano}
\affiliation{Department of Physics and Sezione I.N.F.N., University
of Perugia, Via A. Pascoli, 06123 Italy}
\author{Abolfazl Bayat}
\affiliation{Department of Physics and Astronomy, University College
London, Gower St., London WC1E 6BT, UK}
\author{Sougato Bose}
\affiliation{Department of Physics and Astronomy, University College
London, Gower St., London WC1E 6BT, UK}

\begin{abstract}
We show that, in the gapless Kondo Regime, a single local quench
at one end of a Kondo spin chain induces a fast and long lived
oscillatory dynamics. This quickly establishes a high quality
entanglement between the spins at the opposite ends of the chain.
This entanglement is mediated by the Kondo Cloud, attains a constant
high value independent of the length for large chains, and shows
thermal robustness. In contrast, when the Kondo cloud is absent, e.g. in the gapped dimer regime, only
finite size end to end effects can create some entanglement on a
much longer time-scale for rather short chains. By decoupling one end of the chain during the
dynamics one can distinguish between this end-end effect which
vanishes, and the global Kondo cloud mediated entanglement, which
persists. This quench approach paves the way to detect the elusive
Kondo cloud through the entanglement between two individual spins.
Our results show that non-perturbative cooperative phenomena from
condensed matter may be exploited for quantum information.
\end{abstract}

\date{\today}
\maketitle
\section{INTRODUCTION}
Entanglement in many-body spin systems is
a topic of high interest
 \cite{Amico-RMP,bose-vedral-fazio-osborne,vidal,korepin,Calabrese,Laflorencie-Affleck,Sorensen-Affleck}
 where
``quantum correlations" or long range entanglement between individual spins
is notoriously difficult to achieve. Here we propose to generate such entanglement through a non-perturbative
dynamical mechanism which requires only a \textit{minimal action} on a spin chain,
namely a sudden quench of a single bond. We consider spin chains
with a magnetic impurity \cite{Affleck-Eggert} located at
one end; under certain conditions, the ground state of the chain is such that
the impurity is maximally entangled to the block of spins forming the Kondo Cloud (KC)
\cite{Laflorencie-Affleck,Sorensen-Affleck,Kondous}. We will show
that, during the dynamics induced by the quench, it is the Kondo
cloud which mediates entanglement between the two ending spins of the chain. The generated entanglement is measurable
through two spin correlations and enables a direct detection of
the presence of a KC within the system before quenching.
This is important since, despite several manifestations,
the direct
observation of a KC is still an open challenge \cite{quantum-dot}. Besides
being a quantum informational tool for probing a condensed matter
paradigm, the long range entanglement created by local quenching is
\textit{substantial} and potentially useful for linking separated
quantum registers \cite{bose03-bose-review}.

 Though entanglement in condensed matter systems is typically
very short ranged \cite{bose-vedral-fazio-osborne}, there have
been a few other proposals for long distance entanglement --
however they come at a high price. For example, there are
proposals exploiting weak couplings of two distant spins to a spin
chain \cite{lorenzo1-illuminati,lorenzo2-plenio-li-wojcik}, which
have limited thermal stability or very long time-scale of
entanglement generation. Otherwise, a dynamics has to be induced
by large-scale changes to the Hamiltonian of a system
\cite{Hannu-Hangii}, as opposed to the minimal change required in
our proposal. Additionally, the exploitation of a non-perturbative
cooperative feature of a condensed matter system, such as a Kondo
cloud, has not been addressed so far.

Quantum spin chains allow to investigate non-perturbative
phenomena arising from the presence of impurities
\cite{Affleck-LesHouches08}. For instance, the spin chain Kondo
model
\cite{Laflorencie-Affleck,Sorensen-Affleck,Laflorencie-2008}
arises when a magnetic impurity \cite{Affleck-Eggert} is coupled
to the end of an antiferromagnetic $J_1-J_2$ spin-1/2 chain, where
$J_1 (J_2)$ is the (next) nearest neighbor coupling. The model
supports a crossover from a gapless Kondo Regime (KR) for $J_2<J_2^c=0.2412J_1$ to a gapped Dimerized
Regime (DR) for $J_2>J_2^c$. It is called a Kondo model since the
impurity spin forms, in the KR, an effective singlet
with a the spins inside the KC. Recently, the
entanglement of the ground state of this system has been
characterized by varied means
\cite{Laflorencie-Affleck,Sorensen-Affleck}. The Kondo spin chain
is described by the Hamiltonian
\begin{eqnarray}\label{hamil_NNN}
    H_I&=&J'(J_1\sigma_1.\sigma_2+J_2\sigma_1.\sigma_3)\cr
    &+&J_1\sum_{i=2}^{N-1}\sigma_i.\sigma_{i+1}+J_2\sum_{i=2}^{N-2}\sigma_i.\sigma_{i+2},
\end{eqnarray}
where $\sigma_i=(\sigma_i^x,\sigma_i^y,\sigma_i^z)$ is the vector
of Pauli operators at site $i$, $J_1>0$ $(J_2>0)$ is the anti
ferromagnetic (next) nearest neighbor coupling constants (we always put $J_1=1$), $J'<1$
is the impurity bond strength and $N$ is the number of spins,
which we assume to be even through this paper. $H_I$ is the
initial Hamiltonian of our dynamical scheme. Tuning $J_2$ enables one to shift the spin
chain between the Kondo and DRs.

\section{LONG-RANGE DISTANCE-INDEPENDENT ENTANGLEMENT IN THE KONDO REGIME}
 We consider the {\em
finite} Kondo chain (\ref{hamil_NNN}) in its ground state
$|GS_I\ra$. We, then, pertinently quench the coupling at the
opposite end of the impurity allowing for the dynamics to develop
entanglement. We show that, in the
KR, the entanglement between the two ending spins
oscillates between high peaks with a periodicity determined by
$J'$, while the dynamics is very fast (thereby decoherence hardly
gets time to act) and is robust against thermal fluctuations. In
the dimer phase, the dynamics is much slower, qualitatively
different and, in finite chains, it generates some entanglement due to
unavoidable end to end effects, which are drastically tamed if one "cuts off"
the impurity from the chain during the dynamics. In the KR,
cutting off the impurity has minimal effect on the final
entanglement between the ending spins since, here, the process is
mediated by the cloud.

Entanglement is expected to be very different
\cite{Laflorencie-Affleck,Sorensen-Affleck, Kondous} as $J_2$
crosses $J_2^c$ since, for $J_2<J_2^c$, there is a characteristic
length - the so called Kondo screening length $\xi$ - determined
only by $J'$ through $\xi\propto e^{\alpha/\sqrt{J'}}$ where
$\alpha$ is a constant. In the KR, for any given $J'$
and $N$, $\xi$ determines the size of a domain whose spins are
maximally entangled with the impurity spin sitting at the origin
\cite{Kondous}.

Initially, the system is assumed to be in the ground state
$|GS_I\ra$ of $H_I$. A minimal quench modifies only the
couplings of the $N$th spin by
the amount $J'$ (same as $J'$ in Eq.(\ref{hamil_NNN})) so that
$H_I$ is changed to
\begin{eqnarray}\label{hamil_NNN_quenched}
    H_F&=&J'(J_1\sigma_1.\sigma_2+J_2\sigma_1.\sigma_3+J_1\sigma_{N-1}.\sigma_N+J_2\sigma_{N-2}.\sigma_N)\cr
    &+&J_1\sum_{i=2}^{N-2}\sigma_i.\sigma_{i+1}+J_2\sum_{i=2}^{N-3}\sigma_i.\sigma_{i+2}.
\end{eqnarray}
Since $|GS_I\ra$ is not an eigenstate of $H_F$ it will evolve as
$|\psi(t)\ra=e^{-iH_Ft}|GS_I\ra.$ An entanglement $E(N,t,J')$
between the ending spins emerges as a result of the above
evolution. To compute $E(N,t,J')$, we first obtain the reduced
density matrix $\rho_{1N}(t)=tr_{\hat{1N}}|\psi(t)\ra\la\psi(t)|$
of spins $1$ and $N$ by tracing out the remaining spins from the
state $|\psi(t)\ra$. Then, we evaluate $E(N,t,J')$ in terms of a
measure of entanglement valid for arbitrary mixed states of two
qubits called the concurrence \cite{wootters}.
Entanglement takes its maximum $E_m$ at an optimal time $t_{opt}$ and
an optimal coupling $J'_{opt}$ such that $E_m=E(N,t_{opt},J'_{opt})$.
As we shall see $J'_{opt}$ is not a
perturbation of $J_1$ and $J_2$. If, as expected from scaling in
the KR \cite{Affleck-scaling, Kondous}, the dependence
on $N$ and $t$ can be accounted for by a redefinition of $J'$
(equivalently $\xi$), then $t_{opt}$ and $J'_{opt}$ cannot be
independent quantities. Our numerical analysis shows indeed that,
in the KR, $t_{opt}\propto N$ and that
$J'_{opt}$ yields $\xi=N-2$;
since $\xi\propto e^{\alpha/\sqrt{J'}}$ one gets $t_{opt}\propto N\propto e^{\alpha/\sqrt{J'_{opt}}}$.

For our choice of $J',J_1$ and $J_2$ the spin-chain dynamics is not analytically solvable and one
has to resort to numerical simulations. Recent methods of
many-body simulations allow handling exponentially big Hilbert
spaces with pertinent truncations. Here, for $N>20$, we use the time-step
targeting method, based on the DMRG algorithm introduced in
\cite{whiteDMRG}. For $N<20$, instead, we resort to exact diagonalization.

\begin{figure}
\centering
    \includegraphics[width=8cm,height=5.6cm,angle=0]{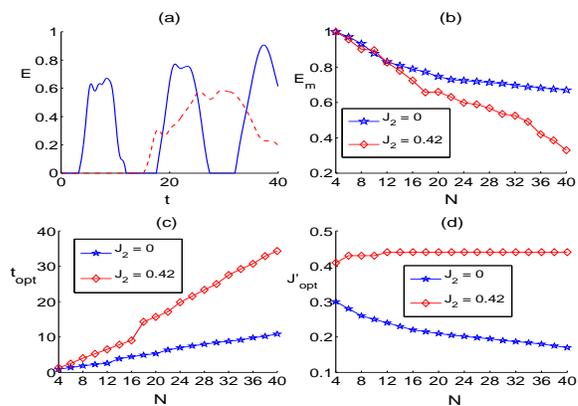}
    \caption{(Color online) Comparing the Kondo ($J_2=0$) and dimer ($J_2=0.42$) regimes.
    a) Entanglement vs. time for $N=30$ with $J'=0.19$ for the KR (solid line) and
     $J'=0.44$ for the DR (dashed line). b) $E_m$ vs. the length $N$.
    c) $t_{opt}$ vs. $N$. d)  $J'_{opt}$ vs. length $N$.}
     \label{Fig1}
\end{figure}

\begin{figure}
\centering
    \includegraphics[width=8cm,height=5.6cm,angle=0]{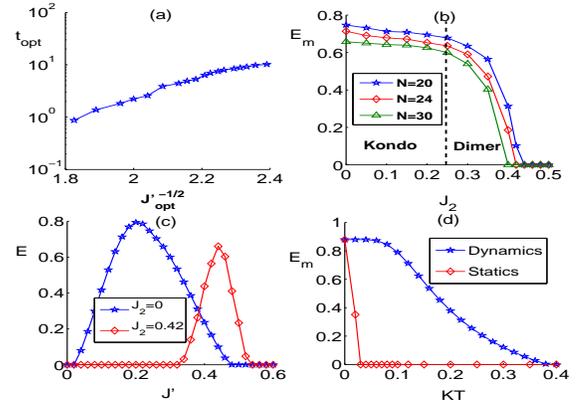}
    \caption{(Color online) a) $t_{opt}$ vs. $1/\sqrt{J'_{opt}}$ in the KR.  b) $E_m$ at $t\propto 1/J'_{opt}$ vs. $J_2$ for chains of different lengths.
    c) Entanglement attained at $t_{opt}$ vs. $J'$ for Kondo and DRs for $N=20$.
    d) $E_m$ vs. temperature after bond quenching (blue line)
    and induced by (static) weak coupling with the rest of the chain (red line) for a chain of $N=10$.}
     \label{Fig2}
\end{figure}

We focus only on the first period of the entanglement evolution
in both regimes, since both decoherence and numerical errors make it unwise to wait for longer times.
 Fig. \ref{Fig1}(a) shows that fast long-lived
(non-decaying) periodic oscillations with a period of $2t_{opt}$ characterize the time evolution in
the KR and that the maximal entanglement is achieved when the impurity coupling $J'$ equals
the value $J'_{opt}$ associated to a KC of size $\xi=N-2$
(the KC generated by the impurity sitting on the left side touches the other side of the chain);
in the DR the dynamics appears
more dispersive and not oscillatory for any $J'$.
In Fig. \ref{Fig1}(b) we plot the maximum of entanglement, $E_m$, induced by bond quenching as a function of
the length $N$: though the entanglement decreases as $N$ increases, its value, in the KR,
stays rather high and becomes almost {\em distance independent} for very long chains; furthermore, as $N$ increases, the
entanglement generated in the KR is significantly bigger than the one
in the DR. Despite its lower value, achieving entanglement in the DR costs
more time, as shown in Fig. \ref{Fig1}(c).
It is also clear from Fig. \ref{Fig1}(c) that $t_{opt}$ increases by $N$ linearly.
Finally, in the KR, $J'_{opt}$ slowly decreases as
$N$ increases while it stays essentially constant in the DR (Fig. \ref{Fig1}(d)).
This is commensurate with the expectation that $t_{opt}\propto e^{\alpha/\sqrt{J'_{opt}}}$ in the KR, while,
in the DR, $t_{opt}$ and $J'_{opt}$ are two independent quantities. The plot in Fig. \ref{Fig2}a shows the
exponential dependence of $t_{opt}$ on $J'^{ -1/2}_{opt}$ realized, in  the KR, for long enough chains.

 How the dynamics creates - even for very long chains of
size $N$- high entanglement oscillations between the ending spins
of the chain in the KR? To understand this, one should
recall that, in the KR, the impurity spin forms an
effective singlet with all the spins inside the cloud
\cite{Kondous} and that, only in this regime, one can always
choose $J'$ so that $\xi$ may be made comparable with $N$; at
variance, in the DR, the impurity in $|GS_I\ra$ picks
out - no matter what the values of $J'$ and $N$ are- only an
individual spin in the chain to form a singlet ( a valence bond)
while the remaining spins form singlets ({\em local dimers}) with
their nearest neighbors \cite{Sorensen-Affleck}. Thus,
 only in the KR, one may use the remarkable resource to select - for any $N$- an initial state
 $|GS_I\ra$ which is free from local excitations: in fact, when $\xi=N-2$  ($J'=J'_{opt}$) there is only a {\emph single entity}, namely the $N$th spin,
 interacting with the impurity-cloud composite.
 As this situation can always be engineered by choosing, for any $N$, $J'=J'_{opt}$, this explains the
distance-independent entanglement in Fig. \ref{Fig1}b. The
proposed scenario provides an intuitive grasp on why, only in the
KR, an optimal entanglement between the ending spins may
emerge from quench dynamics as the result of the interplay between
very few states. At variance, in the DR, the energy
released by quenching is dispersed over the variety of different
quantum modes associated to the local dimers.

To provide a more quantitative analysis, one may
expand $|\psi(t)\ra$ in terms of eigenvectors of $H_F$. By exact diagonalization (up to $N=20$), one finds that, in the
KR, only two eigenstates of $H_F$ (the ground state $|E_1\ra$ and one excited state $|E_2\ra$) predominantly contribute to the dynamics:
\begin{eqnarray}\label{Eigenvectors}
|E_1\ra&=&\alpha_1|\psi^-\ra_{1N}|\phi^-\ra_b+\beta_1(|00\ra_{1N}|\phi^{00}\ra_b\cr
&+&|11\ra_{1N}|\phi^{11}\ra_b
-|\psi^+\ra|\phi^{+}\ra_b)\cr
|E_2\ra&=&\alpha_2|\psi^-\ra_{1N}|\phi^-\ra_b-\beta_2(|00\ra_{1N}|\phi^{00}\ra_b\cr
&+&|11\ra_{1N}|\phi^{11}\ra_b
-|\psi^+\ra|\phi^{+}\ra_b).
\end{eqnarray}
In Eq. (\ref{Eigenvectors}) the first and the last spin are projected onto the singlet ($|\psi^-\ra$) and the
triplets ($|00\ra, |11\ra$ and $|\psi^+\ra$) while the states of all spins in the body of the chain have been
specified by the index $b$.
After a time $t$ the state evolves as
$|\psi(t)\ra=\la E_1|GS_I\ra |E_1\ra+e^{-i\Delta Et}\la E_2|GS_I\ra |E_2\ra+...$,
where $\Delta E$ is the energy separation between the two levels.
One defines $t=t_{opt}$ as the time for which the contribution of
$|\psi^-\ra_{1N}|\phi^-\ra_b$ is most enhanced in $|\psi(t)\ra$ due to
a {\em constructive interference}.

\begin{figure}
\centering
    \includegraphics[width=7cm,height=4.3cm,angle=0]{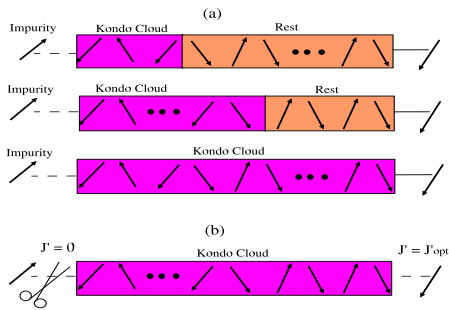}
    \caption{(Color online) a) Different $|GS_I\ra$ for entanglement generation through quench dynamics.
    Top: The ground states with $\xi<N/2$ (no entanglement). Middle: $N/2<\xi<N-2$ (some
    entanglement). Bottom:
    $\xi=N-2$ (optimal entanglement). b) Decoupling the impurity from the chain. }
     \label{Fig3}
\end{figure}

The condition for the onset of constructive interference is
\begin{equation}\label{interference}
|\la E_1|GS_I\ra\beta_1 | \approx |\la E_2|GS_I\ra\beta_2 |,
\end{equation}
so that terms other than $|\psi^-\ra_{1N}|\phi^-\ra_b$ in
$|\psi(t)\ra$ do not contribute at $t=t_{opt}$. When
$J'\rightarrow 1$ (very small cloud) then $|\la E_1|GS_I\ra|
\approx 1$ while $|\la E_2|GS_I\ra| \approx 0$ as the ground state
is hardly changed on quench: thus, constructive interference
between $|E_1\ra$ and $|E_2\ra$ it is impossible. The condition of
Eq. (\ref{interference}) cannot be satisfied also when
$J'\rightarrow 0$, since one has now that $\beta_1 \approx 0$ and
$\beta_2 \approx 1$ (the end spins form a singlet and triplet with
each other in $|E_1\ra$ and $|E_2\ra$ respectively
\cite{lorenzo1-illuminati}). Thus, only for intermediate $J'$
entanglement may peak. When $J'>J_{opt}$ ($\xi<N-2$)- and
particularly for $\xi<N/2$- the entanglement between the ending
spins is frustrated by the existence of local excitations whose
number increases as the size of the cloud gets smaller. In
addition, when $J'<J_{opt}$ ($\xi>N-2$), the KC overtakes
the chain and the $N$th spin is already {\em included in the
cloud} and its tendency is to screen the original impurity as in
$|\psi^-\ra_{1N}|\phi^-\ra_b$ rather than to pair with it to form
a spin one (as in the last three terms of $|E_1\ra$). This makes
$\beta_1$ quite small, and it becomes smaller as the cloud
overtakes the chain and again the condition of Eq.
(\ref{interference}) cannot be fulfilled.
 Consequently, the optimal situation is realized when $J'=J'_{opt}$
 ($\xi=N-2$
 ), i.e just before the cloud overtakes the chain.
Thus, only in the KR, one can convert-for any $N$- the
\emph{useless entanglement} between the impurity spin and the
KC into a \emph{usable entanglement} between the ending
spins of the chain. The emerging long distance entanglement
analyzed in this paper is, indeed, a genuine footprint of the
presence of the KC in $|GS_I\ra$ (Fig. \ref{Fig3}(a)).

In Fig. \ref{Fig2}b we plot- for both regimes- the entanglement reached after waiting for a time interval of the
order of $1/J'_{opt}$. One notices that, for $J_2>J_2^c$, the entanglement peak decreases sensibly and goes
to zero rather soon. The plot of the maximal entanglement vs. $J'$ is given in Fig. \ref{Fig2}c: here one sees that, in
the KR, the entanglement rises from zero already at very small values of $J'$. This is expected since,
in the KR, to a small $J'$ is associated a large cloud containing the impurity sitting on the left side.

\begin{figure}
\centering
    \includegraphics[width=6cm,height=4.5cm,angle=0]{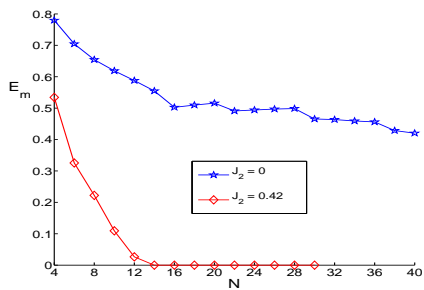}
    \caption{(Color online) $E_m$ vs. $N$ after decoupling the first impurity.}
     \label{Fig4}
\end{figure}

The essential role of the KC in the entanglement
generation is further probed if one let evolve the ground state
$|GS_I\ra$ with a doubly quenched Hamiltonian obtained from
(\ref{hamil_NNN_quenched}) by isolating (i.e., putting $J'=0$) the
left hand side impurity while keeping
fixed to $J'_{opt}$ the bond connected to the last spin (see Fig.
\ref{Fig3}b). This forbids the dynamical build up of that
``portion" of the total entanglement which is only due to end
to end effects. In Fig. \ref{Fig4} we have plotted the  $E_m$ vs. $N$ after double quenching in
both regimes. Fig. \ref{Fig4} shows that entanglement in the dimer
phase collapses already when $N>12$ while
it stays \emph{unexpectedly} high - and almost independent on $N$ - in the KR; the existing entanglement
of the KC with the impurity \cite{Kondous} is dynamically swapped over to the last spin.

 Note that a long
distance singlet between the end spins may be realized in a ground state when those spins
are very weakly coupled ($J'=\epsilon/\sqrt{N}<<1/\sqrt{N}$)
to a spin chain \cite{lorenzo1-illuminati}. This static approach to generate entanglement relies on couplings which are so weak that
they can merely be regarded as perturbations. Such entanglement
is not robust against thermal fluctuations
due to the smallness of the gap ($\propto J'^2=\epsilon^2/N$) between the ground state and a triplet state between the end spins.
On the other hand our approach enables to generate entanglement dynamically even for $J'$ as high as $J'_{opt}\approx 1/(\log{N})^2$.
Even when temperature is increased, the entanglement is not seriously disrupted till $K_BT$ exceeds the Kondo temperature
($\propto 1/\xi=1/(N-2)$) after which the KC does not form.
As a result, while in the dynamical approach $K_BT< 1/(N-2)$, in the static approach one has $K_BT<\epsilon^2/N$: thus, the long distance
entanglement generated through the dynamical approach is thermally
more stable. For instance, for $\epsilon\sim 10^{-1}$, our dynamical approach is robust for temperatures 100 times
higher than those required for the static approach.
In Fig. \ref{Fig2}d, we plot $E_m$- as obtained in both
approaches- vs. temperature for $N=10$. In the
static approach, the ground state is replaced by the thermal
state, whereas in our dynamic approach it is the initial state
which is taken to be the relevant thermal state. We ignore
thermalization and relaxation during dynamics since the dynamical
time scale, set by $t_{opt}$, is fast enough (this
is also an advantage over slow dynamical schemes
with weak couplings \cite{lorenzo2-plenio-li-wojcik}).

There are systems where the Kondo
 cloud mediated long distance entanglement
 may be observed such as spin chains in ion traps \cite{Cirac-Porras-Wunderlich-Porras-Nature-2008},
 with trapped electrons \cite{Marzoli-Tombesi}, in chains of $P$ donors in $Si$ \cite{Kane} and Josephson
 chains with impurities \cite{glala}.

\section{Conclusions} 

We have shown that substantial {\em long range distance independent} entanglement
 can be engineered by a non-perturbative quenching of a {\em single} bond in a Kondo spin chain.
 This is the first example where a \textit{minimal local} action on a spin chain dynamically creates long range entanglement.
 In contrast to all known schemes for entanglement between individual spins in spin chains, here the entanglement attains a constant
 value for long chains rather than decaying with distance. We showed that, in the KR, the entanglement between the ending spins
   is mediated by the the
 KC.
 As the coupling is non-perturbative in strength ($J'_{opt}\approx 1/(\log{N})^2$), the entanglement generation
 is both fast and thermally robust.
 In the KR, the entanglement is periodic
 in time with a period of $2t_{opt}$ -- this is a curious instance of exciting
 a long lived regular oscillation in a gapless system.
  Both, the
 long distance entanglement mediated by the cloud and the periodic
 dynamics, provide a clear
 signature of the existence of the KC in a quantum system with
 impurities. These features are absent when the KC is absent, e.g., in the gapped DR. Beyond the
 spin chain based implementations, in principle, in double quantum
 dots one may detect the length of the KC by inducing
 and probing the entanglement oscillations between the dots. Our
 analysis evidences that interesting applications to quantum information
 may arise exploiting relevant non-perturbative cooperative phenomena of condensed matter physics.

Discussions with I. Affleck, A. Briggs, P. Calabrese, L. Campos
Venuti and M. B. Plenio are warmly acknowledged. AB and SB (PS) thank(s) the University of Perugia (UCL)
for hospitality. AB and PS thank the G.G.I. for
hospitality and INFN for partial support. PS was supported by the
ESF Network INSTANS . AB and SB are supported by the EPSRC, and SB is also supported by the QIPIRC
(GR/S82176 /01), the Royal Society and the Wolfson Foundation.

\end{document}